\documentclass[aps,prl,reprint,showpacs,groupedaddress]{revtex4-1}
\usepackage{subfigure}
\usepackage{amsmath,graphicx,amssymb,float}
\usepackage{multirow,color}
\usepackage{url}
\usepackage{natbib}

\usepackage{bm,verbatim}

\renewcommand{\vec}[1]{\bm{#1}}
\newcommand{\etal}{\textit{et al.}}

\newcommand{\eps}{\varepsilon}

\newcommand{\beq}{\begin{equation}}
\newcommand{\eeq}{\end{equation}}

\begin{document}

\title{Sudden relaminarisation and lifetimes in forced isotropic turbulence}

\author{Moritz F. Linkmann}
\email{m.linkmann@ed.ac.uk}
\author{Alexander Morozov}
\affiliation{SUPA, School of Physics and Astronomy, University of Edinburgh, JCMB, King's Buildings, Peter Guthrie Tait Road, EH9 3FD, Edinburgh, United Kingdom}

\begin{abstract}
We demonstrate an unexpected connection between isotropic turbulence and wall-bounded shear flows. We perform direct numerical simulations of isotropic turbulence forced at large scales at moderate Reynolds numbers and observe sudden transitions from chaotic dynamics to a spatially simple flow, analogous to the laminar state in wall bounded shear flows. 
We find that the survival probabilities of turbulence are exponential and the typical lifetimes increase super-exponentially 
with the Reynolds number. Our results suggest that both isotropic turbulence and wall-bounded shear flows qualitatively share the same phase-space dynamics.
\end{abstract}


\maketitle

Recent years have seen significant advances in our understanding of the
transition to turbulence in wall-bounded shear flows. In simple geometries,
 like flow in a pipe or a channel, close to the transition threshold, a finite-amplitude
 perturbation develops into a localised turbulent patch (a `puff' in pipe flow)
 that exists as an independent entity \cite{Darbyshire95,Hof2004,Nishi08,Duguet2009,Mellibovsky2009,Schneider2010,Duguet2012,Avila2013}. Experiments \cite{Hof06,Avila11} 
 and numerical simulations \cite{Schmiegel97,Faisst04,Avila10} have shown that the
 localised patches of turbulence can spontaneously disappear (relaminarise) or
 split into two. The rates of these competing processes depend strongly on the
 Reynolds number: at relatively low Reynolds numbers it is much more probable for
 a puff to decay than to split, 
while the opposite is true at higher Reynolds numbers.
 The point where the two probabilities are equal marks the transition to a sustained 
turbulence \cite{Avila11}, and the turbulence below this threshold may
consist of long-lived chaotic transients \cite{Crutchfield88}.
The transition to turbulence in wall-bounded flows is thus intimately 
related to the process of relaminarisation, where turbulent dynamics
suddenly collapse to a much simpler, typically linearly stable, laminar 
state. Such events have been explained by dynamical systems theory
as the escape from a chaotic saddle in state space with a 
constant (time independent) rate of escape \cite{Brosa89,Ott02,Eckhardt07,Eckhardt08,Bruno2008}. 
At higher Reynolds numbers, spatially local relaminarisation attempts 
\cite{XiGraham2010,XiGraham2012} can be the source of intermittency in turbulent flows.

In contrast, stationary isotropic turbulence, that can be thought of as
a turbulent flow far away from boundaries \cite{Monin75b}, 
is believed to exhibit much simpler dynamics: its motion is turbulent for 
all Reynolds numbers and there is no actual transition.
In this Letter we report an unexpected connection
between these two fields. We perform direct numerical simulations (DNS) of stationary
isotropic turbulence at low Reynolds numbers and observe sudden breakdowns of
the turbulent dynamics in favour of a much simpler state. Similar observations have 
been made in connection to symmetry-breaking  
in isotropic turbulence \cite{McComb15b} and in magnetohydrodynamic 
flows subject to electrical forcing \cite{Dallas14a}.
We study the nature of this process and show that it is analogous 
to the relaminarisation events in wall-bounded parallel shear flows. 
We find that forced isotropic turbulence at relatively low Reynolds 
numbers is transient and the rate of its collapse is constant in time,
resulting in exponentially distributed lifetimes of the turbulent state similar to pipe 
\cite{Hof06,Eckhardt07,Avila10,Avila11} 
and plane Couette flow \cite{Schmiegel97,Bottin98,Shi13}.

We perform direct numerical simulations of the incompressible Navier-Stokes equations (NSE)
\begin{align}
\label{eq:nse}
\partial_t \vec{u}&= - \nabla P -\vec{u}\cdot \nabla\vec{u} + \nu \Delta \vec{u} + \vec{f}, \\ 
\label{eq:incompr}
&\nabla \cdot \vec{u} = 0,
\end{align}
where $\vec{u}$ denotes the velocity field, $\vec{f}$ is an external force, 
$\nu$ is the kinematic viscosity,
$P$ is the pressure, and we set the density to unity. These equations were solved
numerically using the standard fully de-aliased pseudospectral method \cite{Yoffe12} on a 3D periodic domain of length 
$L_{box}=2\pi$ with the smallest wavenumber being 
$k_{min}=2\pi/L_{box} =1$. All simulations are well-resolved, using $32^3$ collocation points and
satisfying $k_{max}\eta \geqslant 1.82$, where $\eta$ denotes the Kolmogorov dissipation scale. 

The system is forced at large scales by a negative damping $\vec{f}$ defined as
\begin{align}
 \hat{\vec{f}}(\vec{k},t) &=
      (\varepsilon_W/2 E_f) \hat{\vec{u}}(\vec{k},t) \quad
\text{for} \quad  0 < \lvert\vec{k}\rvert < k_f ; \nonumber \\
  &= 0   \quad \textrm{otherwise}.
\label{eq:forcing}
\end{align}
Here, $\hat{\vec{f}}(\vec{k},t) $ is the Fourier transform of the forcing,
$\hat{\vec{u}}(\vec{k},t)$ is the Fourier transform of the velocity
field $\vec{u}(\vec{x},t)$, $E_f$ is the total energy contained in the forcing
band, and $k_f = 2.5$ is the highest wavenumber forced. Normalizing 
the energy input by $E_f$ ensures that the energy injection rate is 
$\varepsilon_W =\textrm{constant}$; here we choose $\eps_W=0.1$.
This forcing provides an energy input that does not
prefer any particular direction and has a complicated, time-dependent spatial
profile; note that $k_f =2.5$ corresponds to  
80 possible wavevectors and thus 80 different velocity field modes are being forced.  
It is commonly used in numerical investigations of 
homogeneous isotropic turbulence 
\cite{Jimenez93,McComb01a,Yamazaki02,Kaneda03,McComb03,McComb14b,McComb14c},
the prime example being the series of high-resolution simulations of Kaneda 
\etal~\cite{Kaneda06}.

The initial conditions for the velocity with a prescribed energy spectrum are constructed by assigning a Gaussian random vector
to each point in space. The resulting field is subsequently Fourier-transformed and rescaled according to 
the desired energy spectrum in the form
\begin{equation}
E(k)=0.001702\,k^4 e^{-2(k/5)^2}.
\end{equation} 
Further details of the numerical method, validation and 
benchmarking of the code can be found in \cite{Yoffe12}.

The simulations are evolved for $1271$ initial large-eddy
turnover times $t_0=L/U$, where $U$ denotes 
the initial rms velocity and $L$ is the initial integral length scale; $t_0=0.78$ in simulation units.
The parameter that is varied in our simulations is the viscosity $\nu$, and we
present the results in terms of a system-scale Reynolds number $Re=L_{box}^{4/3}\eps_W^{1/3}/\nu$
that is changed from $53.80$ to $97.82$ 
for different simulations; in each individual run, $Re$ is kept constant during the whole simulation.
The corresponding values of the Taylor-Reynolds number
and further simulation details are given in Supplementary Material~\footnote{See supplementary material for details.}.

\begin{figure}[tbp]
 \begin{center}
   \includegraphics[width=\columnwidth]{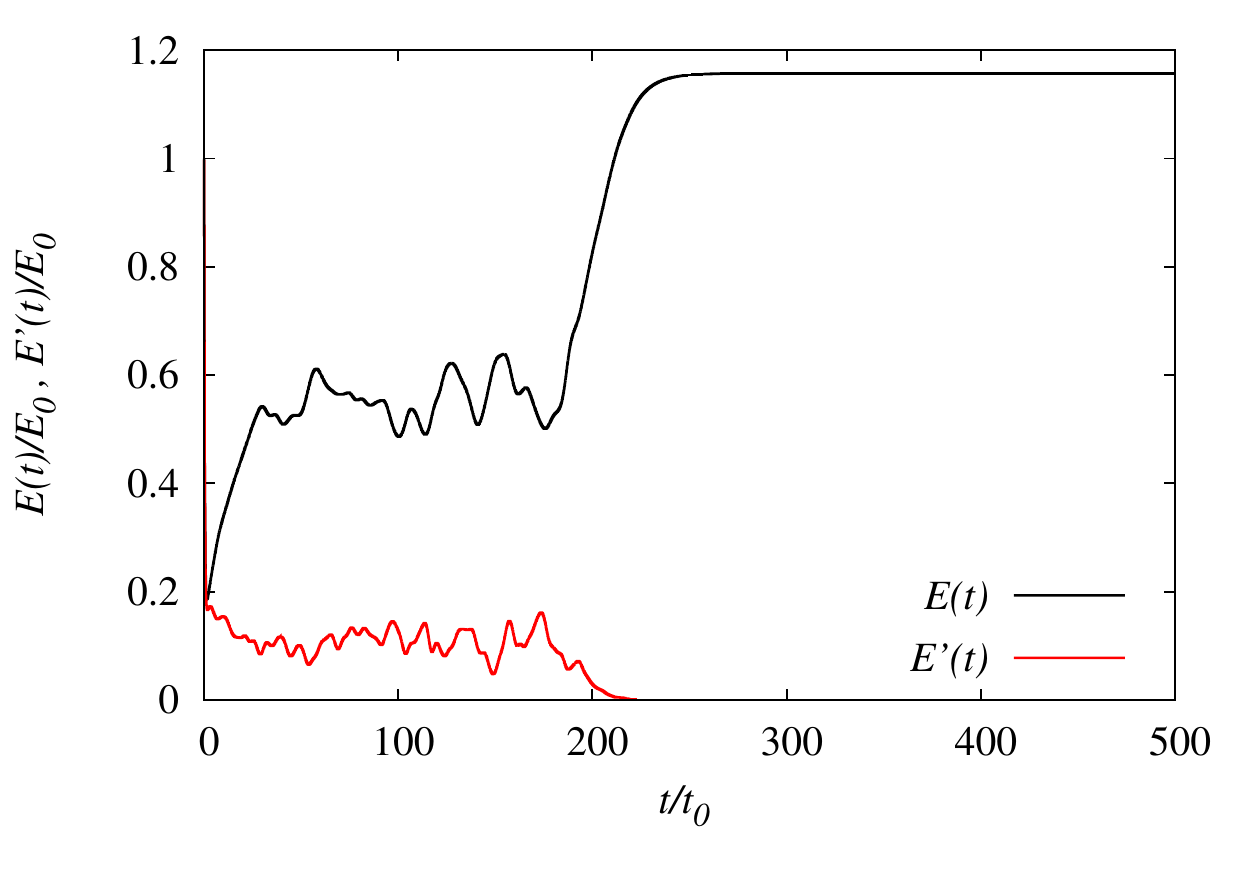}
 \end{center}
 \caption{(Color online) Time evolution of the total energy $E(t)$ and the energy content of the small scales $E'(t)$ 
for $Re=76.86$ normalised by the initial energy $E_0$. Time is given in units of initial 
large eddy turnover time $t_0=L/U$, where $U$ is the initial rms velocity and $L$ the initial integral scale. 
The point around $t/t_0\approx 240$ when $E'(t)$ vanishes and the total energy becomes constant marks the onset of the self-organised state as discussed in the main text.}
\label{fig:hit3d}
\end{figure}

As mentioned above, the form of the forcing term that we employ here, Eq.\eqref{eq:forcing}, is routinely used
in DNS of isotropic turbulence as its complicated spatial form would seem to guarantee that the system
is turbulent at any Reynolds number larger than unity. Indeed, even at sufficiently low Reynolds numbers, 
we observe that our simulations reach a turbulent stationary state, where the energy injection is balanced by the average dissipation and there is motion at all lengthscales. Surprisingly, however, after staying in this steady state for a long time, the system exhibits a transition to a different state, as shown, for example, in Fig.~\ref{fig:hit3d} for $Re=76.86$. There we plot the total energy of the system, $E(t)=\int_{k_{min}}^{k_{max}} dk \ E(k,t)$ and the energy content of small scales, $E'(t)=\int_{k>k_{min}}^{k_{max}} dk \ E(k,t)$, as a function of time; the energy content at a particular lengthscale is $E(k,t) = \lvert \hat{\vec{u}}(k,t) \rvert^2/2$ and the largest scale in the system corresponds to $k_{min}=2\pi/L_{box}=1$.

As Fig.~\ref{fig:hit3d} demonstrates, turbulent dynamics persists until about $t/t_0 \approx 240$. After that, the total energy becomes constant and the small-scale fluctuations
in the kinetic energy produced by the characteristic turbulent cascade process 
suddenly disappear. This implies that for $t/t_0 > 240$ 
the kinetic energy is confined to the largest scale of the system and 
no nonlinear transfer exciting the smaller scales takes place. 
The system thus transitions from a turbulent to a large-scale `laminar' state. 

The existence of such a state can be understood if one considers a model velocity field with $u_x \sim \cos(y)$ and all other components of the velocity being zero. This flow profile is similar to a simple shear flow: it satisfies the incompressibility condition, it does not produce any pressure gradient in the system, and the non-linear term vanishes exactly for this profile. It is, therefore, an exact solution of the equations of motion, Eqs.\eqref{eq:nse}-\eqref{eq:incompr}, with its magnitude being set by the injection rate $\eps_W$ and the kinematic viscosity $\nu$. In general, one can construct many exact solutions of the Navier-Stokes equations with $k=1$, similar to the model profile discussed above, for which the non-linear term vanishes. What is surprising, however, is that this self-organised large-scale state is dynamically connected to the isotropic turbulence at sufficiently low Reynolds numbers.

When the system selects this self-organised state, it stays there for as long as our simulations continue. 
Together with the fact that this state is dynamically selected by the system, it seems to imply that this 
state is linearly stable. In order to further probe this statement, we have performed exploratory simulations 
where we have perturbed the self-organised state with random perturbations and observed their evolution. 
For sufficiently small amplitude of the perturbations, simulations always returned to the self-organised state, 
while for larger perturbations the system became turbulent, 
as shown in Fig.~2 of the Supplementary Material \cite{Note1}. 
Therefore the simple state reported here has the same property as the laminar 
state in many wall-bounded parallel shear flows (cf. the Hagen-Poiseuille profile in 
pipe flow \cite{Drazin04}): it is a linearly stable simple exact solution that can be destabilised by a finite-amplitude perturbation. 

Next, we observe that at a fixed Reynolds number, the time of self-organisation ($t/t_0\approx 240$ in the example above) strongly depends on the initial conditions. 
We explore this variability systematically by starting $100$ runs with different initial conditions for a fixed value of $Re$. In each simulation, we monitor the time-evolution of the total energy $E(t)$ and the dissipation rate $\eps(t) = 2\nu \int_{k_{min}}^{k_{max}} dk\ k^2E(k,t)$. 
In order to identify the moment when the turbulent dynamics collapses onto the self-organised state, we employ a criterium that is based on the observation that 
since the kinetic energy in the self-organized state is confined to modes with
$k=1$, the asymptotic value $E_{\infty}$ for 
all individual runs in a given ensemble (at a given $Re$) 
can be calculated 
from the energy input rate  $\varepsilon_W$ and $\nu$. 
For statistically stationary flows the energy input rate $\eps_W$
must equal the dissipation rate $\eps$, 
and we obtain for the total energy of the self-organised state
\begin{equation}
\label{eq:E_B}
E_{\infty}  =\frac{\varepsilon_W}{2\nu}=\mbox{ constant}.
\end{equation}
Our data confirms that in every simulation,
the total energy eventually reaches the asymptotic value $E_{\infty}$, and the self-organisation time
can be defined as the time when $E(t)=E_{\infty}$. 
We have checked that when we define the relaminarisation moment as the time when the dissipation rate equals the input rate without any fluctuations, we obtain identical results.

\begin{figure}[!tbp]
 \begin{center}
   \includegraphics[width=\columnwidth]{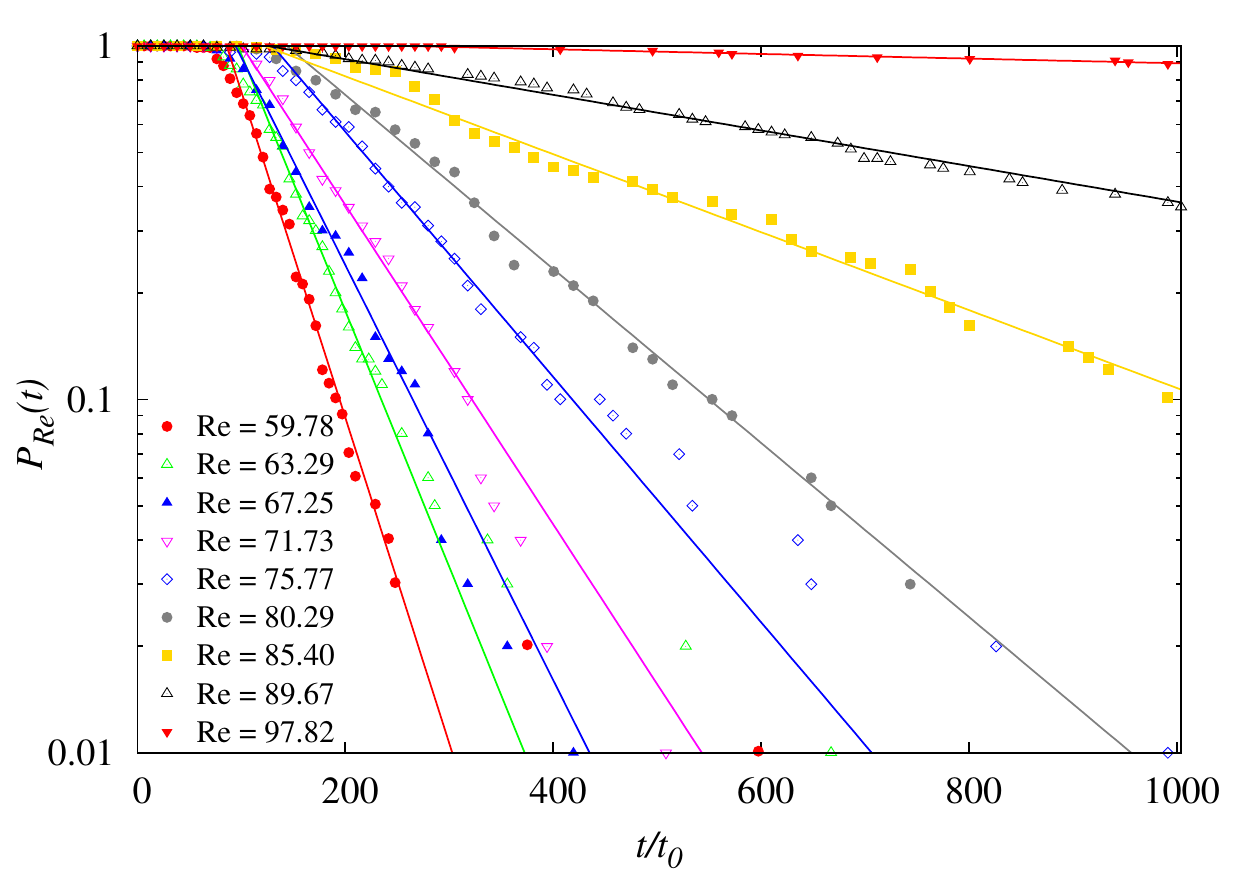}
 \end{center}
 \caption{(Color online) Survival probability as a function of the dimensionless time $t/t_0$ from the beginning of a simulation.}
 \label{fig:survival_prob}
\end{figure}

We quantify the variability of the self-organisation times by introducing a survival probability
$P_{Re}(t)$ that at a given $Re$ gives the probability that the system is still turbulent at time $t$, 
having started in a turbulent state at time $t=0$. For each $t$, we estimate this probability
by dividing the number of runs that are still turbulent after time $t$ by the total number of runs performed at
this Reynolds number.
The resulting survival probabilities are shown in Fig.\ref{fig:survival_prob} for a range
of $Re$. We find that after some initial lag time during which the system has evolved from the initial
condition into the turbulent state, the survival probability follows a simple exponential
law
\begin{equation}
 \label{eq:survival_prob}
 P_{Re}(t)\sim\exp(-t/\tau(Re)),
\end{equation}
where $\tau(Re)$ is the typical lifetime of turbulence that only depends on the Reynolds number.
The exponential form of the survival probability suggests that the process is memoryless, i.e. at each time
the rate of relaminarisation is constant and does not depend on the previous dynamics of the system.
This behaviour is identical to what was observed in wall-bounded shear flows, such as pipe \cite{Hof06,Eckhardt07,Hof08,Avila10,Avila11} or plane 
Couette flow \cite{Schmiegel97,Bottin98,Shi13}.
There, it was attributed to the escape from a chaotic saddle 
associated with relaminarisation of localised turbulence \cite{Eckhardt07,Hof08,Eckhardt08}.

In order to verify that our results do not depend on the size of the simulation box,
one ensemble of 100 runs was created using a larger simulation box with 
$L_{box} = 4 \pi$. The collapse of turbulence is also observed in these runs and leads to an  
exponential survival probability with the same characteristic lifetime as 
a reference dataset at $L_{box} = 2 \pi$ \cite{Note1}.

\begin{figure}[!tbp]
 \begin{center}
   \includegraphics[width=\columnwidth]{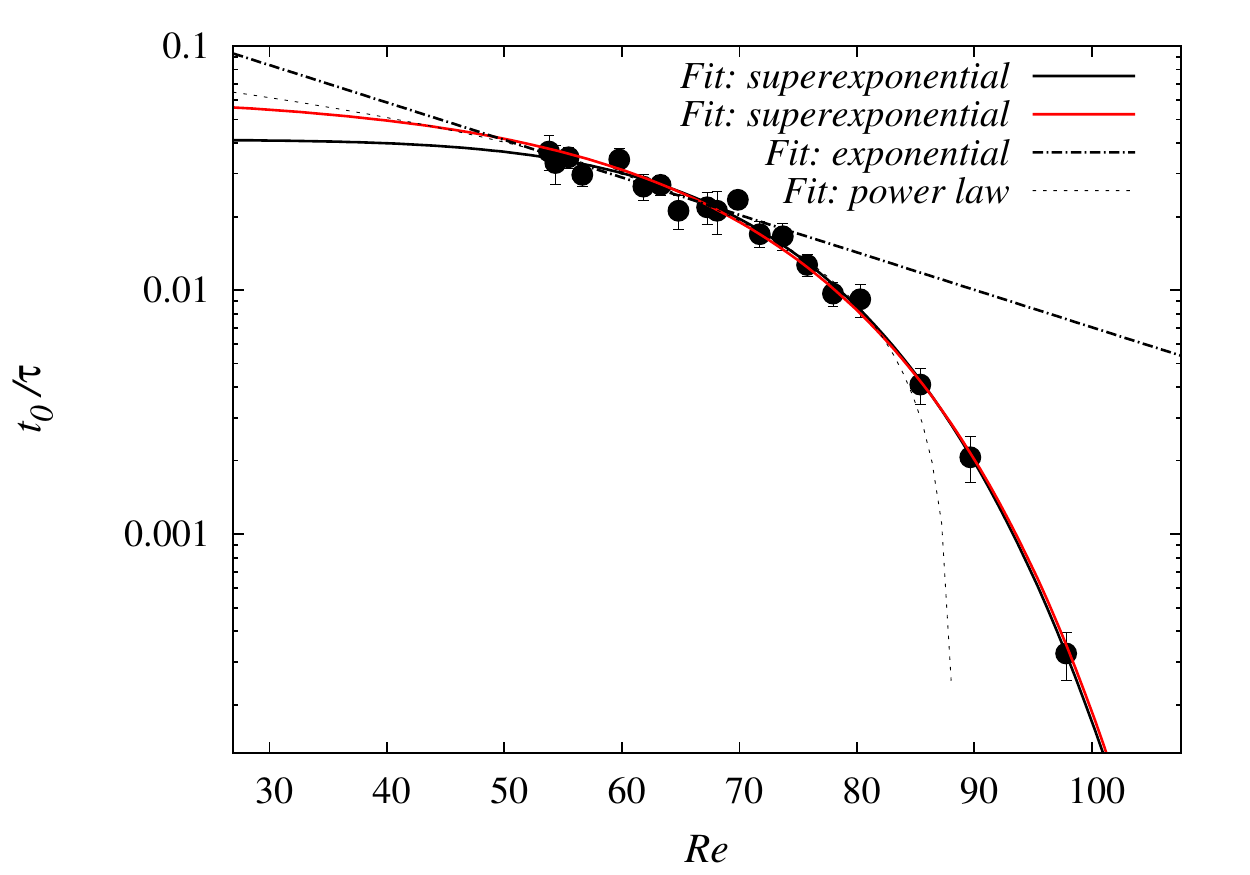}
 \end{center}
 \caption{(Color online) Reynolds number dependence of the escape rate 
          $t_0/\tau$. The red (grey) line is a two-parameter fit of the expression
         $t_0/\tau(Re)= 0.064\exp(-\exp[a+b Re])$, the black line a 
         two-parameter fit of the expression
         $t_0/\tau(Re)= \exp[a'-(b' Re)^{5.6}])$, the dash-dotted line a fit
         of an exponential and the faint dotted line a fit of a linear dependence
         of $t_0/\tau$ on Reynolds number. 
}
 \label{fig:escape_Re}
\end{figure}

The characteristic lifetime $\tau$ is obtained at each Reynolds
number from fitting the survival probabilities to Eq.\eqref{eq:survival_prob}, see solid lines in Fig.~\ref{fig:survival_prob}. 
We observe a steep increase in $\tau$ with increasing Reynolds number 
as shown in Fig.~\ref{fig:escape_Re}. To find the functional form $\tau=\tau(Re)$, we 
fit the observed lifetime to various model expressions. First we consider a power law with an exponent $n<0$ in the form $\tau \sim (Re_c-Re)^n$ that would suggest a divergence of the lifetime at some critical Reynolds number $Re_c$. We find that it is not compatible with the data for any value of $n$; Fig.~\ref{fig:escape_Re} shows an example with $n=-1$. The same applies to an exponential increase of $\tau$ with $Re$. However, we find that a super-exponential scaling in the form
\begin{equation}
\label{eq:tau_Re}
\frac{\tau(Re)}{t_0} = c \exp\left[\exp(a+b Re)\right]
\end{equation}
is compatible with our data for a fixed amplitude $c=15.63$ and $a=-3.48 \pm 0.51$, $b=0.052 \pm 0.005$, see Fig.~\ref{fig:escape_Re}. Once again, this conclusion parallels the super-exponential scaling of the lifetimes in wall-bounded shear flows \cite{Hof06,Hof08,Eckhardt08}. Further support for this scaling is provided in the Supplementary Material \cite{Note1}.

We note that the super-exponential law Eq.\eqref{eq:tau_Re} is not 
the only possible form that produces an acceptable fit to the data. Another super-exponential
dependence, $\tau(Re)/t_0= \exp[-a'+(b' Re)^{5.6}])$ with 
$a'=-3.18 \pm 0.14$ and $b'=0.0136 \pm 0.0003$ also gives a good agreement with the dataset, 
as can be seen in Fig.~\ref{fig:escape_Re}.  

The results presented in this Letter show that there is a surprising analogy between the behaviour of the isotropic turbulence forced at a large scale and wall-bounded shear flows at low Reynolds numbers. We observe that there is a spontaneous transition from turbulence to a spatially-simple state, which we have identified here, and this ``laminar" state is linearly stable but can be destabilised by a finite-amplitude perturbation. The turbulent-laminar transition is abrupt and memoryless, and the associated survival probability is exponential in time, cf. \cite{Ott02,Hof06,Eckhardt07,Hof08,Avila10,Avila11}. The turbulent lifetimes do not diverge with an increase in $Re$, but instead grow super-exponentially, cf. \cite{Hof08,Goldenfeld2010}. This analogy implies that the phenomena of the transition to turbulence in wall-bounded shear flows and forced isotropic turbulence, typically thought of as a high-$Re$ phenomenon away from boundaries, are dynamically similar and can be understood within the same theoretical framework. 
As recent research suggests, the transition to turbulence in shear flows belongs to the directed percolation universality 
class \cite{Goldenfeld2011,HofPreprint2015,Shi13,ShihPreprint15}, and we argue that the same might be valid for forced isotropic turbulence.

\begin{figure}[!tbp]
\begin{center}
\includegraphics[width=\columnwidth]{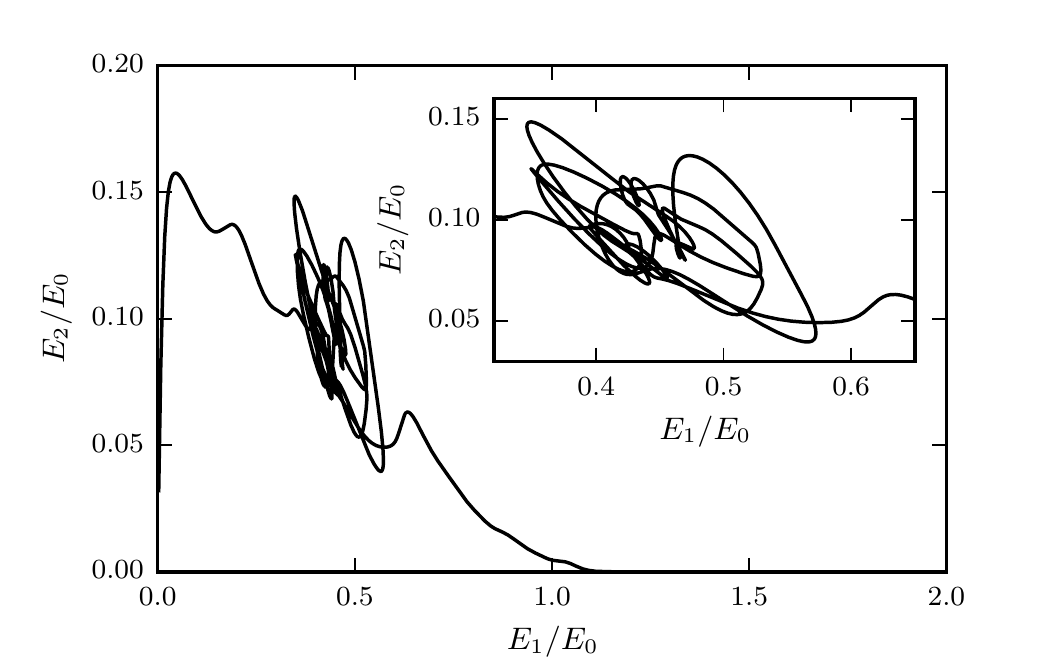}
\end{center}
\caption{
Phase portrait $E_2$ vs $E_1$ for $Re=76.86$. Each point corresponds to a particular moment in time. All energies are scaled with the initial total kinetic energy $E_0$. Inset: Zoom of the turbulent region of the main graph showing that the dynamics is organised by several points in phase space suggestive of unstable exact solutions.}
\label{fig:portrait}
\end{figure}

The phase space of turbulent wall-bounded shear flows is organised by exact solutions and periodic orbits of the Navier-Stokes equations \cite{Faisst03,Cvitanovic10} and the relaminarisation events are associated with a sudden escape from this part of the phase space \cite{Ott02}. Since we observe the same phenomenology, we speculate that the phase space of the forced isotropic turbulence should also be organised by coherent structures (exact solutions and periodic orbits). In Fig.~\ref{fig:portrait} we plot the energy content in the $k=2$ mode vs the energy in the $k=1$ mode for a run at $Re=76.86$. Each point there corresponds to a particular moment in time and the dynamics proceeds from left to right, until the system relaminarises (i.e. $E_1=E_\infty$ and $E_2=0$). We observe that the dynamics revolves around several points in phase space that are very suggestive of exact unstable solutions \cite{Cvitanovic10}. Identification of these coherent states will be the subject of future work.

This work also suggests that the type of forcing employed here is well-suited for DNS of isotropic turbulence with a view of creating an artificial, simpler system whose dynamics still resemble more complicated real physical systems, such as 
shear flows. Other types of forcing, notably various forms of stochastic forcing, are routinely used but might not have the phenomenological similarities with the transition to turbulence in wall-bounded shear flows. Recent results on self-organisation in magnetohydrodynamic flows, for example, 
demonstrate that introducing random phases in the forcing term precludes the formation
of a large-scale flow \cite{Dallas14a}. We argue that Eq.\eqref{eq:forcing} gives in fact a better approximation to naturally occurring turbulence than an explicitly stochastic (and thus more costly) forcing.

Our results provide new potential targets for turbulence control, since we have shown
that there is a stable large-scale state hidden in what appears to be isotropic turbulence. 
A particular choice of an additional external force may be sufficient to push the system into the 
basin of attraction of this stable state.  

We would like to thank Bruno Eckhardt, W.~David McComb, 
Arjun Berera and Samuel Yoffe for helpful discussions,
and Bernardas Jankauskas and Richard Ho for initial help with simulations. 
This work has made use of the resources provided by the 
Edinburgh Compute and Data Facility ({\tt http://www.ecdf.ed.ac.uk}). 
M.F.L. and A.M. acknowledge support from the 
UK Engineering and Physical Sciences Research Council (EP/K503034/1 and EP/I004262/1).

\bibliography{refs}	 

\begin{thebibliography}{45}%
\makeatletter
\providecommand \@ifxundefined [1]{%
 \@ifx{#1\undefined}
}%
\providecommand \@ifnum [1]{%
 \ifnum #1\expandafter \@firstoftwo
 \else \expandafter \@secondoftwo
 \fi
}%
\providecommand \@ifx [1]{%
 \ifx #1\expandafter \@firstoftwo
 \else \expandafter \@secondoftwo
 \fi
}%
\providecommand \natexlab [1]{#1}%
\providecommand \enquote  [1]{``#1''}%
\providecommand \bibnamefont  [1]{#1}%
\providecommand \bibfnamefont [1]{#1}%
\providecommand \citenamefont [1]{#1}%
\providecommand \href@noop [0]{\@secondoftwo}%
\providecommand \href [0]{\begingroup \@sanitize@url \@href}%
\providecommand \@href[1]{\@@startlink{#1}\@@href}%
\providecommand \@@href[1]{\endgroup#1\@@endlink}%
\providecommand \@sanitize@url [0]{\catcode `\\12\catcode `\$12\catcode
  `\&12\catcode `\#12\catcode `\^12\catcode `\_12\catcode `\%12\relax}%
\providecommand \@@startlink[1]{}%
\providecommand \@@endlink[0]{}%
\providecommand \url  [0]{\begingroup\@sanitize@url \@url }%
\providecommand \@url [1]{\endgroup\@href {#1}{\urlprefix }}%
\providecommand \urlprefix  [0]{URL }%
\providecommand \Eprint [0]{\href }%
\providecommand \doibase [0]{http://dx.doi.org/}%
\providecommand \selectlanguage [0]{\@gobble}%
\providecommand \bibinfo  [0]{\@secondoftwo}%
\providecommand \bibfield  [0]{\@secondoftwo}%
\providecommand \translation [1]{[#1]}%
\providecommand \BibitemOpen [0]{}%
\providecommand \bibitemStop [0]{}%
\providecommand \bibitemNoStop [0]{.\EOS\space}%
\providecommand \EOS [0]{\spacefactor3000\relax}%
\providecommand \BibitemShut  [1]{\csname bibitem#1\endcsname}%
\let\auto@bib@innerbib\@empty
\bibitem [{\citenamefont {Darbyshire}\ and\ \citenamefont
  {Mullin}(1995)}]{Darbyshire95}%
  \BibitemOpen
  \bibfield  {author} {\bibinfo {author} {\bibfnamefont {A.~G.}\ \bibnamefont
  {Darbyshire}}\ and\ \bibinfo {author} {\bibfnamefont {T.}~\bibnamefont
  {Mullin}},\ }\href@noop {} {\bibfield  {journal} {\bibinfo  {journal} {J.
  Fluid Mech.}\ }\textbf {\bibinfo {volume} {289}},\ \bibinfo {pages} {83}
  (\bibinfo {year} {1995})}\BibitemShut {NoStop}%
\bibitem [{\citenamefont {Hof}\ \emph {et~al.}(2004)\citenamefont {Hof},
  \citenamefont {van Doorne}, \citenamefont {Westerweel}, \citenamefont
  {Nieuwstadt}, \citenamefont {Faisst}, \citenamefont {Eckhardt}, \citenamefont
  {Wedin}, \citenamefont {Kerswell},\ and\ \citenamefont {Waleffe}}]{Hof2004}%
  \BibitemOpen
  \bibfield  {author} {\bibinfo {author} {\bibfnamefont {B.}~\bibnamefont
  {Hof}}, \bibinfo {author} {\bibfnamefont {C.~W.~H.}\ \bibnamefont {van
  Doorne}}, \bibinfo {author} {\bibfnamefont {J.}~\bibnamefont {Westerweel}},
  \bibinfo {author} {\bibfnamefont {F.~T.~M.}\ \bibnamefont {Nieuwstadt}},
  \bibinfo {author} {\bibfnamefont {H.}~\bibnamefont {Faisst}}, \bibinfo
  {author} {\bibfnamefont {B.}~\bibnamefont {Eckhardt}}, \bibinfo {author}
  {\bibfnamefont {H.}~\bibnamefont {Wedin}}, \bibinfo {author} {\bibfnamefont
  {R.~R.}\ \bibnamefont {Kerswell}}, \ and\ \bibinfo {author} {\bibfnamefont
  {F.}~\bibnamefont {Waleffe}},\ }\href@noop {} {\bibfield  {journal} {\bibinfo
   {journal} {Science}\ }\textbf {\bibinfo {volume} {305}},\ \bibinfo {pages}
  {1594} (\bibinfo {year} {2004})}\BibitemShut {NoStop}%
\bibitem [{\citenamefont {Nishi}\ \emph {et~al.}(2008)\citenamefont {Nishi},
  \citenamefont {\"{U}nsal}, \citenamefont {Durst},\ and\ \citenamefont
  {Biswas}}]{Nishi08}%
  \BibitemOpen
  \bibfield  {author} {\bibinfo {author} {\bibfnamefont {M.}~\bibnamefont
  {Nishi}}, \bibinfo {author} {\bibfnamefont {B.}~\bibnamefont {\"{U}nsal}},
  \bibinfo {author} {\bibfnamefont {F.}~\bibnamefont {Durst}}, \ and\ \bibinfo
  {author} {\bibfnamefont {G.}~\bibnamefont {Biswas}},\ }\href@noop {}
  {\bibfield  {journal} {\bibinfo  {journal} {J. Fluid Mech.}\ }\textbf
  {\bibinfo {volume} {614}},\ \bibinfo {pages} {425} (\bibinfo {year}
  {2008})}\BibitemShut {NoStop}%
\bibitem [{\citenamefont {Duguet}\ \emph {et~al.}(2009)\citenamefont {Duguet},
  \citenamefont {Schlatter},\ and\ \citenamefont {Henningson}}]{Duguet2009}%
  \BibitemOpen
  \bibfield  {author} {\bibinfo {author} {\bibfnamefont {Y.}~\bibnamefont
  {Duguet}}, \bibinfo {author} {\bibfnamefont {P.}~\bibnamefont {Schlatter}}, \
  and\ \bibinfo {author} {\bibfnamefont {D.~S.}\ \bibnamefont {Henningson}},\
  }\href@noop {} {\bibfield  {journal} {\bibinfo  {journal} {Phys. Fluids}\
  }\textbf {\bibinfo {volume} {21}},\ \bibinfo {pages} {111701} (\bibinfo
  {year} {2009})}\BibitemShut {NoStop}%
\bibitem [{\citenamefont {Mellibovsky}\ \emph {et~al.}(2009)\citenamefont
  {Mellibovsky}, \citenamefont {Meseguer}, \citenamefont {Schneider},\ and\
  \citenamefont {Eckhardt}}]{Mellibovsky2009}%
  \BibitemOpen
  \bibfield  {author} {\bibinfo {author} {\bibfnamefont {F.}~\bibnamefont
  {Mellibovsky}}, \bibinfo {author} {\bibfnamefont {A.}~\bibnamefont
  {Meseguer}}, \bibinfo {author} {\bibfnamefont {T.~M.}\ \bibnamefont
  {Schneider}}, \ and\ \bibinfo {author} {\bibfnamefont {B.}~\bibnamefont
  {Eckhardt}},\ }\href@noop {} {\bibfield  {journal} {\bibinfo  {journal}
  {Phys. Rev. Lett.}\ }\textbf {\bibinfo {volume} {103}},\ \bibinfo {pages}
  {054502} (\bibinfo {year} {2009})}\BibitemShut {NoStop}%
\bibitem [{\citenamefont {Schneider}\ \emph {et~al.}(2010)\citenamefont
  {Schneider}, \citenamefont {Marinc},\ and\ \citenamefont
  {Eckhardt}}]{Schneider2010}%
  \BibitemOpen
  \bibfield  {author} {\bibinfo {author} {\bibfnamefont {T.~M.}\ \bibnamefont
  {Schneider}}, \bibinfo {author} {\bibfnamefont {D.}~\bibnamefont {Marinc}}, \
  and\ \bibinfo {author} {\bibfnamefont {B.}~\bibnamefont {Eckhardt}},\
  }\href@noop {} {\bibfield  {journal} {\bibinfo  {journal} {Journal of Fluid
  Mechanics}\ }\textbf {\bibinfo {volume} {646}},\ \bibinfo {pages} {441}
  (\bibinfo {year} {2010})}\BibitemShut {NoStop}%
\bibitem [{\citenamefont {Duguet}\ \emph {et~al.}(2012)\citenamefont {Duguet},
  \citenamefont {Schlatter}, \citenamefont {Henningson},\ and\ \citenamefont
  {Eckhardt}}]{Duguet2012}%
  \BibitemOpen
  \bibfield  {author} {\bibinfo {author} {\bibfnamefont {Y.}~\bibnamefont
  {Duguet}}, \bibinfo {author} {\bibfnamefont {P.}~\bibnamefont {Schlatter}},
  \bibinfo {author} {\bibfnamefont {D.~S.}\ \bibnamefont {Henningson}}, \ and\
  \bibinfo {author} {\bibfnamefont {B.}~\bibnamefont {Eckhardt}},\ }\href@noop
  {} {\bibfield  {journal} {\bibinfo  {journal} {Phys. Rev. Lett.}\ }\textbf
  {\bibinfo {volume} {108}},\ \bibinfo {pages} {044501} (\bibinfo {year}
  {2012})}\BibitemShut {NoStop}%
\bibitem [{\citenamefont {Avila}\ \emph {et~al.}(2013)\citenamefont {Avila},
  \citenamefont {Mellibovsky}, \citenamefont {Roland},\ and\ \citenamefont
  {Hof}}]{Avila2013}%
  \BibitemOpen
  \bibfield  {author} {\bibinfo {author} {\bibfnamefont {M.}~\bibnamefont
  {Avila}}, \bibinfo {author} {\bibfnamefont {F.}~\bibnamefont {Mellibovsky}},
  \bibinfo {author} {\bibfnamefont {N.}~\bibnamefont {Roland}}, \ and\ \bibinfo
  {author} {\bibfnamefont {B.}~\bibnamefont {Hof}},\ }\href@noop {} {\bibfield
  {journal} {\bibinfo  {journal} {Phys. Rev. Lett.}\ }\textbf {\bibinfo
  {volume} {110}},\ \bibinfo {pages} {224502} (\bibinfo {year}
  {2013})}\BibitemShut {NoStop}%
\bibitem [{\citenamefont {Hof}\ \emph {et~al.}(2006)\citenamefont {Hof},
  \citenamefont {Westerweel}, \citenamefont {Schneider},\ and\ \citenamefont
  {Eckhardt}}]{Hof06}%
  \BibitemOpen
  \bibfield  {author} {\bibinfo {author} {\bibfnamefont {B.}~\bibnamefont
  {Hof}}, \bibinfo {author} {\bibfnamefont {J.}~\bibnamefont {Westerweel}},
  \bibinfo {author} {\bibfnamefont {T.~M.}\ \bibnamefont {Schneider}}, \ and\
  \bibinfo {author} {\bibfnamefont {B.}~\bibnamefont {Eckhardt}},\ }\href@noop
  {} {\bibfield  {journal} {\bibinfo  {journal} {Nature}\ }\textbf {\bibinfo
  {volume} {443}},\ \bibinfo {pages} {59} (\bibinfo {year} {2006})}\BibitemShut
  {NoStop}%
\bibitem [{\citenamefont {Avila}\ \emph {et~al.}(2011)\citenamefont {Avila},
  \citenamefont {Moxey}, \citenamefont {{de Lozar}}, \citenamefont {Avila},
  \citenamefont {Barkley},\ and\ \citenamefont {Hof}}]{Avila11}%
  \BibitemOpen
  \bibfield  {author} {\bibinfo {author} {\bibfnamefont {K.}~\bibnamefont
  {Avila}}, \bibinfo {author} {\bibfnamefont {D.}~\bibnamefont {Moxey}},
  \bibinfo {author} {\bibfnamefont {A.}~\bibnamefont {{de Lozar}}}, \bibinfo
  {author} {\bibfnamefont {M.}~\bibnamefont {Avila}}, \bibinfo {author}
  {\bibfnamefont {D.}~\bibnamefont {Barkley}}, \ and\ \bibinfo {author}
  {\bibfnamefont {B.}~\bibnamefont {Hof}},\ }\href@noop {} {\bibfield
  {journal} {\bibinfo  {journal} {Science}\ }\textbf {\bibinfo {volume}
  {333}},\ \bibinfo {pages} {192} (\bibinfo {year} {2011})}\BibitemShut
  {NoStop}%
\bibitem [{\citenamefont {Schmiegel}\ and\ \citenamefont
  {Eckhardt}(1997)}]{Schmiegel97}%
  \BibitemOpen
  \bibfield  {author} {\bibinfo {author} {\bibfnamefont {A.}~\bibnamefont
  {Schmiegel}}\ and\ \bibinfo {author} {\bibfnamefont {B.}~\bibnamefont
  {Eckhardt}},\ }\href@noop {} {\bibfield  {journal} {\bibinfo  {journal}
  {Phys. Rev. Lett.}\ }\textbf {\bibinfo {volume} {79}},\ \bibinfo {pages}
  {5250} (\bibinfo {year} {1997})}\BibitemShut {NoStop}%
\bibitem [{\citenamefont {Faisst}\ and\ \citenamefont
  {Eckhardt}(2004)}]{Faisst04}%
  \BibitemOpen
  \bibfield  {author} {\bibinfo {author} {\bibfnamefont {H.}~\bibnamefont
  {Faisst}}\ and\ \bibinfo {author} {\bibfnamefont {B.}~\bibnamefont
  {Eckhardt}},\ }\href@noop {} {\bibfield  {journal} {\bibinfo  {journal} {J.
  Fluid Mech.}\ }\textbf {\bibinfo {volume} {504}},\ \bibinfo {pages} {343}
  (\bibinfo {year} {2004})}\BibitemShut {NoStop}%
\bibitem [{\citenamefont {Avila}\ \emph {et~al.}(2010)\citenamefont {Avila},
  \citenamefont {Willis},\ and\ \citenamefont {Hof}}]{Avila10}%
  \BibitemOpen
  \bibfield  {author} {\bibinfo {author} {\bibfnamefont {M.}~\bibnamefont
  {Avila}}, \bibinfo {author} {\bibfnamefont {A.~P.}\ \bibnamefont {Willis}}, \
  and\ \bibinfo {author} {\bibfnamefont {B.}~\bibnamefont {Hof}},\ }\href@noop
  {} {\bibfield  {journal} {\bibinfo  {journal} {J. Fluid Mech.}\ }\textbf
  {\bibinfo {volume} {646}},\ \bibinfo {pages} {127} (\bibinfo {year}
  {2010})}\BibitemShut {NoStop}%
\bibitem [{\citenamefont {Crutchfield}\ and\ \citenamefont
  {Kaneko}(1988)}]{Crutchfield88}%
  \BibitemOpen
  \bibfield  {author} {\bibinfo {author} {\bibfnamefont {J.~P.}\ \bibnamefont
  {Crutchfield}}\ and\ \bibinfo {author} {\bibfnamefont {K.}~\bibnamefont
  {Kaneko}},\ }\href@noop {} {\bibfield  {journal} {\bibinfo  {journal} {Phys.
  Rev. Lett.}\ }\textbf {\bibinfo {volume} {60}},\ \bibinfo {pages} {2715}
  (\bibinfo {year} {1988})}\BibitemShut {NoStop}%
\bibitem [{\citenamefont {Brosa}(1989)}]{Brosa89}%
  \BibitemOpen
  \bibfield  {author} {\bibinfo {author} {\bibfnamefont {U.}~\bibnamefont
  {Brosa}},\ }\href@noop {} {\bibfield  {journal} {\bibinfo  {journal} {J.
  Stat. Phys.}\ }\textbf {\bibinfo {volume} {55}},\ \bibinfo {pages} {1301}
  (\bibinfo {year} {1989})}\BibitemShut {NoStop}%
\bibitem [{\citenamefont {Ott}(2002)}]{Ott02}%
  \BibitemOpen
  \bibfield  {author} {\bibinfo {author} {\bibfnamefont {E.}~\bibnamefont
  {Ott}},\ }\href@noop {} {\emph {\bibinfo {title} {Chaos in Dynamical
  Systems}}},\ \bibinfo {edition} {2nd}\ ed.\ (\bibinfo  {publisher} {Cambridge
  University Press, Cambridge},\ \bibinfo {year} {2002})\BibitemShut {NoStop}%
\bibitem [{\citenamefont {Eckhardt}\ \emph {et~al.}(2007)\citenamefont
  {Eckhardt}, \citenamefont {Schneider}, \citenamefont {Hof},\ and\
  \citenamefont {Westerweel}}]{Eckhardt07}%
  \BibitemOpen
  \bibfield  {author} {\bibinfo {author} {\bibfnamefont {B.}~\bibnamefont
  {Eckhardt}}, \bibinfo {author} {\bibfnamefont {T.~M.}\ \bibnamefont
  {Schneider}}, \bibinfo {author} {\bibfnamefont {B.}~\bibnamefont {Hof}}, \
  and\ \bibinfo {author} {\bibfnamefont {J.}~\bibnamefont {Westerweel}},\
  }\href@noop {} {\bibfield  {journal} {\bibinfo  {journal} {Ann. Rev. Fluid
  Mech.}\ }\textbf {\bibinfo {volume} {39}},\ \bibinfo {pages} {447} (\bibinfo
  {year} {2007})}\BibitemShut {NoStop}%
\bibitem [{\citenamefont {Eckhardt}\ \emph {et~al.}(2008)\citenamefont
  {Eckhardt}, \citenamefont {Faisst}, \citenamefont {Schmiegel},\ and\
  \citenamefont {Schneider}}]{Eckhardt08}%
  \BibitemOpen
  \bibfield  {author} {\bibinfo {author} {\bibfnamefont {B.}~\bibnamefont
  {Eckhardt}}, \bibinfo {author} {\bibfnamefont {H.}~\bibnamefont {Faisst}},
  \bibinfo {author} {\bibfnamefont {A.}~\bibnamefont {Schmiegel}}, \ and\
  \bibinfo {author} {\bibfnamefont {T.~M.}\ \bibnamefont {Schneider}},\
  }\href@noop {} {\bibfield  {journal} {\bibinfo  {journal} {Phil. Trans. R.
  Soc. A}\ }\textbf {\bibinfo {volume} {366}},\ \bibinfo {pages} {1297}
  (\bibinfo {year} {2008})}\BibitemShut {NoStop}%
\bibitem [{\citenamefont {Eckhardt}\ and\ \citenamefont
  {Schneider}(2008)}]{Bruno2008}%
  \BibitemOpen
  \bibfield  {author} {\bibinfo {author} {\bibfnamefont {B.}~\bibnamefont
  {Eckhardt}}\ and\ \bibinfo {author} {\bibfnamefont {T.~M.}\ \bibnamefont
  {Schneider}},\ }\href@noop {} {\bibfield  {journal} {\bibinfo  {journal}
  {Eur. Phys. J. B}\ }\textbf {\bibinfo {volume} {64}},\ \bibinfo {pages} {457}
  (\bibinfo {year} {2008})}\BibitemShut {NoStop}%
\bibitem [{\citenamefont {Xi}\ and\ \citenamefont
  {Graham}(2010)}]{XiGraham2010}%
  \BibitemOpen
  \bibfield  {author} {\bibinfo {author} {\bibfnamefont {L.}~\bibnamefont
  {Xi}}\ and\ \bibinfo {author} {\bibfnamefont {M.~D.}\ \bibnamefont
  {Graham}},\ }\href@noop {} {\bibfield  {journal} {\bibinfo  {journal} {Phys.
  Rev. Lett.}\ }\textbf {\bibinfo {volume} {104}},\ \bibinfo {pages} {218301}
  (\bibinfo {year} {2010})}\BibitemShut {NoStop}%
\bibitem [{\citenamefont {Xi}\ and\ \citenamefont
  {Graham}(2012)}]{XiGraham2012}%
  \BibitemOpen
  \bibfield  {author} {\bibinfo {author} {\bibfnamefont {L.}~\bibnamefont
  {Xi}}\ and\ \bibinfo {author} {\bibfnamefont {M.~D.}\ \bibnamefont
  {Graham}},\ }\href@noop {} {\bibfield  {journal} {\bibinfo  {journal} {J.
  Fluid Mech.}\ }\textbf {\bibinfo {volume} {693}},\ \bibinfo {pages} {433}
  (\bibinfo {year} {2012})}\BibitemShut {NoStop}%
\bibitem [{\citenamefont {Monin}\ and\ \citenamefont
  {Yaglom}(1975)}]{Monin75b}%
  \BibitemOpen
  \bibfield  {author} {\bibinfo {author} {\bibfnamefont {A.~S.}\ \bibnamefont
  {Monin}}\ and\ \bibinfo {author} {\bibfnamefont {A.~M.}\ \bibnamefont
  {Yaglom}},\ }\href@noop {} {\emph {\bibinfo {title} {{Statistical Fluid
  Mechanics}}}},\ \bibinfo {edition} {2nd}\ ed.\ (\bibinfo  {publisher} {MIT
  Press},\ \bibinfo {year} {1975})\BibitemShut {NoStop}%
\bibitem [{\citenamefont {McComb}\ \emph
  {et~al.}(2015{\natexlab{a}})\citenamefont {McComb}, \citenamefont {Linkmann},
  \citenamefont {Berera}, \citenamefont {Yoffe},\ and\ \citenamefont
  {Jankauskas}}]{McComb15b}%
  \BibitemOpen
  \bibfield  {author} {\bibinfo {author} {\bibfnamefont {W.~D.}\ \bibnamefont
  {McComb}}, \bibinfo {author} {\bibfnamefont {M.~F.}\ \bibnamefont
  {Linkmann}}, \bibinfo {author} {\bibfnamefont {A.}~\bibnamefont {Berera}},
  \bibinfo {author} {\bibfnamefont {S.~R.}\ \bibnamefont {Yoffe}}, \ and\
  \bibinfo {author} {\bibfnamefont {B.}~\bibnamefont {Jankauskas}},\
  }\href@noop {} {\bibfield  {journal} {\bibinfo  {journal} {J. Phys. A: Math.
  Theor.}\ }\textbf {\bibinfo {volume} {48}},\ \bibinfo {pages} {25FT01}
  (\bibinfo {year} {2015}{\natexlab{a}})}\BibitemShut {NoStop}%
\bibitem [{\citenamefont {Dallas}\ and\ \citenamefont
  {Alexakis}(2015)}]{Dallas14a}%
  \BibitemOpen
  \bibfield  {author} {\bibinfo {author} {\bibfnamefont {V.}~\bibnamefont
  {Dallas}}\ and\ \bibinfo {author} {\bibfnamefont {A.}~\bibnamefont
  {Alexakis}},\ }\href@noop {} {\bibfield  {journal} {\bibinfo  {journal}
  {Phys. Fluids}\ }\textbf {\bibinfo {volume} {27}},\ \bibinfo {pages} {045105}
  (\bibinfo {year} {2015})}\BibitemShut {NoStop}%
\bibitem [{\citenamefont {Bottin}\ and\ \citenamefont
  {Chat\'{e}}(1998)}]{Bottin98}%
  \BibitemOpen
  \bibfield  {author} {\bibinfo {author} {\bibfnamefont {S.}~\bibnamefont
  {Bottin}}\ and\ \bibinfo {author} {\bibfnamefont {H.}~\bibnamefont
  {Chat\'{e}}},\ }\href@noop {} {\bibfield  {journal} {\bibinfo  {journal}
  {Eur. Phys. J. B}\ }\textbf {\bibinfo {volume} {6}},\ \bibinfo {pages} {143}
  (\bibinfo {year} {1998})}\BibitemShut {NoStop}%
\bibitem [{\citenamefont {Shi}\ \emph {et~al.}(2013)\citenamefont {Shi},
  \citenamefont {Avila},\ and\ \citenamefont {Hof}}]{Shi13}%
  \BibitemOpen
  \bibfield  {author} {\bibinfo {author} {\bibfnamefont {L.}~\bibnamefont
  {Shi}}, \bibinfo {author} {\bibfnamefont {M.}~\bibnamefont {Avila}}, \ and\
  \bibinfo {author} {\bibfnamefont {B.}~\bibnamefont {Hof}},\ }\href@noop {}
  {\bibfield  {journal} {\bibinfo  {journal} {Phys. Rev. Lett.}\ }\textbf
  {\bibinfo {volume} {110}},\ \bibinfo {pages} {204502} (\bibinfo {year}
  {2013})}\BibitemShut {NoStop}%
\bibitem [{\citenamefont {Yoffe}(2012)}]{Yoffe12}%
  \BibitemOpen
  \bibfield  {author} {\bibinfo {author} {\bibfnamefont {S.~R.}\ \bibnamefont
  {Yoffe}},\ }\emph {\bibinfo {title} {{Investigation of the transfer and
  dissipation of energy in isotropic turbulence}}},\ \href@noop {} {Ph.D.
  thesis},\ \bibinfo  {school} {University of Edinburgh} (\bibinfo {year}
  {2012}),\ \bibinfo {note} {arXiv:1306.3408}\BibitemShut {NoStop}%
\bibitem [{\citenamefont {Jim{\'e}nez}\ \emph {et~al.}(1993)\citenamefont
  {Jim{\'e}nez}, \citenamefont {Wray}, \citenamefont {Saffman},\ and\
  \citenamefont {Rogallo}}]{Jimenez93}%
  \BibitemOpen
  \bibfield  {author} {\bibinfo {author} {\bibfnamefont {J.}~\bibnamefont
  {Jim{\'e}nez}}, \bibinfo {author} {\bibfnamefont {A.~A.}\ \bibnamefont
  {Wray}}, \bibinfo {author} {\bibfnamefont {P.~G.}\ \bibnamefont {Saffman}}, \
  and\ \bibinfo {author} {\bibfnamefont {R.~S.}\ \bibnamefont {Rogallo}},\
  }\href@noop {} {\bibfield  {journal} {\bibinfo  {journal} {J. Fluid Mech.}\
  }\textbf {\bibinfo {volume} {255}},\ \bibinfo {pages} {65} (\bibinfo {year}
  {1993})}\BibitemShut {NoStop}%
\bibitem [{\citenamefont {McComb}\ \emph {et~al.}(2001)\citenamefont {McComb},
  \citenamefont {Hunter},\ and\ \citenamefont {Johnston}}]{McComb01a}%
  \BibitemOpen
  \bibfield  {author} {\bibinfo {author} {\bibfnamefont {W.~D.}\ \bibnamefont
  {McComb}}, \bibinfo {author} {\bibfnamefont {A.}~\bibnamefont {Hunter}}, \
  and\ \bibinfo {author} {\bibfnamefont {C.}~\bibnamefont {Johnston}},\
  }\href@noop {} {\bibfield  {journal} {\bibinfo  {journal} {Phys. Fluids}\
  }\textbf {\bibinfo {volume} {13}},\ \bibinfo {pages} {2030} (\bibinfo {year}
  {2001})}\BibitemShut {NoStop}%
\bibitem [{\citenamefont {Yamazaki}\ \emph {et~al.}(2002)\citenamefont
  {Yamazaki}, \citenamefont {Ishihara},\ and\ \citenamefont
  {Kaneda}}]{Yamazaki02}%
  \BibitemOpen
  \bibfield  {author} {\bibinfo {author} {\bibfnamefont {Y.}~\bibnamefont
  {Yamazaki}}, \bibinfo {author} {\bibfnamefont {T.}~\bibnamefont {Ishihara}},
  \ and\ \bibinfo {author} {\bibfnamefont {Y.}~\bibnamefont {Kaneda}},\
  }\href@noop {} {\bibfield  {journal} {\bibinfo  {journal} {J. Phys. Soc.
  Jap.}\ }\textbf {\bibinfo {volume} {71}},\ \bibinfo {pages} {777} (\bibinfo
  {year} {2002})}\BibitemShut {NoStop}%
\bibitem [{\citenamefont {Kaneda}\ \emph {et~al.}(2003)\citenamefont {Kaneda},
  \citenamefont {Ishihara}, \citenamefont {Yokokawa}, \citenamefont {Itakura},\
  and\ \citenamefont {Uno}}]{Kaneda03}%
  \BibitemOpen
  \bibfield  {author} {\bibinfo {author} {\bibfnamefont {Y.}~\bibnamefont
  {Kaneda}}, \bibinfo {author} {\bibfnamefont {T.}~\bibnamefont {Ishihara}},
  \bibinfo {author} {\bibfnamefont {M.}~\bibnamefont {Yokokawa}}, \bibinfo
  {author} {\bibfnamefont {K.}~\bibnamefont {Itakura}}, \ and\ \bibinfo
  {author} {\bibfnamefont {A.}~\bibnamefont {Uno}},\ }\href@noop {} {\bibfield
  {journal} {\bibinfo  {journal} {Phys. Fluids}\ }\textbf {\bibinfo {volume}
  {15}},\ \bibinfo {pages} {L21} (\bibinfo {year} {2003})}\BibitemShut
  {NoStop}%
\bibitem [{\citenamefont {McComb}\ and\ \citenamefont
  {Quinn}(2003)}]{McComb03}%
  \BibitemOpen
  \bibfield  {author} {\bibinfo {author} {\bibfnamefont {W.~D.}\ \bibnamefont
  {McComb}}\ and\ \bibinfo {author} {\bibfnamefont {A.~P.}\ \bibnamefont
  {Quinn}},\ }\href@noop {} {\bibfield  {journal} {\bibinfo  {journal} {Physica
  A}\ }\textbf {\bibinfo {volume} {317}},\ \bibinfo {pages} {487} (\bibinfo
  {year} {2003})}\BibitemShut {NoStop}%
\bibitem [{\citenamefont {McComb}\ \emph {et~al.}(2014)\citenamefont {McComb},
  \citenamefont {Yoffe}, \citenamefont {Linkmann},\ and\ \citenamefont
  {Berera}}]{McComb14b}%
  \BibitemOpen
  \bibfield  {author} {\bibinfo {author} {\bibfnamefont {W.~D.}\ \bibnamefont
  {McComb}}, \bibinfo {author} {\bibfnamefont {S.~R.}\ \bibnamefont {Yoffe}},
  \bibinfo {author} {\bibfnamefont {M.~F.}\ \bibnamefont {Linkmann}}, \ and\
  \bibinfo {author} {\bibfnamefont {A.}~\bibnamefont {Berera}},\ }\href@noop {}
  {\bibfield  {journal} {\bibinfo  {journal} {Phys. Rev. E}\ }\textbf {\bibinfo
  {volume} {90}},\ \bibinfo {pages} {053010} (\bibinfo {year}
  {2014})}\BibitemShut {NoStop}%
\bibitem [{\citenamefont {McComb}\ \emph
  {et~al.}(2015{\natexlab{b}})\citenamefont {McComb}, \citenamefont {Berera},
  \citenamefont {Yoffe},\ and\ \citenamefont {Linkmann}}]{McComb14c}%
  \BibitemOpen
  \bibfield  {author} {\bibinfo {author} {\bibfnamefont {W.~D.}\ \bibnamefont
  {McComb}}, \bibinfo {author} {\bibfnamefont {A.}~\bibnamefont {Berera}},
  \bibinfo {author} {\bibfnamefont {S.~R.}\ \bibnamefont {Yoffe}}, \ and\
  \bibinfo {author} {\bibfnamefont {M.~F.}\ \bibnamefont {Linkmann}},\
  }\href@noop {} {\bibfield  {journal} {\bibinfo  {journal} {Phys. Rev. E}\
  }\textbf {\bibinfo {volume} {91}},\ \bibinfo {pages} {043013} (\bibinfo
  {year} {2015}{\natexlab{b}})}\BibitemShut {NoStop}%
\bibitem [{\citenamefont {Kaneda}\ and\ \citenamefont
  {Ishihara}(2006)}]{Kaneda06}%
  \BibitemOpen
  \bibfield  {author} {\bibinfo {author} {\bibfnamefont {Y.}~\bibnamefont
  {Kaneda}}\ and\ \bibinfo {author} {\bibfnamefont {T.}~\bibnamefont
  {Ishihara}},\ }\href@noop {} {\bibfield  {journal} {\bibinfo  {journal}
  {Journal of Turbulence}\ }\textbf {\bibinfo {volume} {7}},\ \bibinfo {pages}
  {1} (\bibinfo {year} {2006})}\BibitemShut {NoStop}%
\bibitem [{Note1()}]{Note1}%
  \BibitemOpen
  \bibinfo {note} {See supplementary material for details.}\BibitemShut {Stop}%
\bibitem [{\citenamefont {Drazin}\ and\ \citenamefont {Reid}(2004)}]{Drazin04}%
  \BibitemOpen
  \bibfield  {author} {\bibinfo {author} {\bibfnamefont {P.}~\bibnamefont
  {Drazin}}\ and\ \bibinfo {author} {\bibfnamefont {W.}~\bibnamefont {Reid}},\
  }\href@noop {} {\emph {\bibinfo {title} {Hydrodynamic Stability}}}\ (\bibinfo
   {publisher} {Cambridge University Press, Cambridge},\ \bibinfo {year}
  {2004})\BibitemShut {NoStop}%
\bibitem [{\citenamefont {Hof}\ \emph {et~al.}(2008)\citenamefont {Hof},
  \citenamefont {{de Lozar}}, \citenamefont {Kuik},\ and\ \citenamefont
  {Westerweel}}]{Hof08}%
  \BibitemOpen
  \bibfield  {author} {\bibinfo {author} {\bibfnamefont {B.}~\bibnamefont
  {Hof}}, \bibinfo {author} {\bibfnamefont {A.}~\bibnamefont {{de Lozar}}},
  \bibinfo {author} {\bibfnamefont {D.~J.}\ \bibnamefont {Kuik}}, \ and\
  \bibinfo {author} {\bibfnamefont {J.}~\bibnamefont {Westerweel}},\
  }\href@noop {} {\bibfield  {journal} {\bibinfo  {journal} {Phys. Rev. Lett.}\
  }\textbf {\bibinfo {volume} {101}},\ \bibinfo {pages} {214501} (\bibinfo
  {year} {2008})}\BibitemShut {NoStop}%
\bibitem [{\citenamefont {Goldenfeld}\ \emph {et~al.}(2010)\citenamefont
  {Goldenfeld}, \citenamefont {Guttenberg},\ and\ \citenamefont
  {Gioia}}]{Goldenfeld2010}%
  \BibitemOpen
  \bibfield  {author} {\bibinfo {author} {\bibfnamefont {N.}~\bibnamefont
  {Goldenfeld}}, \bibinfo {author} {\bibfnamefont {N.}~\bibnamefont
  {Guttenberg}}, \ and\ \bibinfo {author} {\bibfnamefont {G.}~\bibnamefont
  {Gioia}},\ }\href@noop {} {\bibfield  {journal} {\bibinfo  {journal} {Phys.
  Rev. E}\ }\textbf {\bibinfo {volume} {81}},\ \bibinfo {pages} {035304}
  (\bibinfo {year} {2010})}\BibitemShut {NoStop}%
\bibitem [{\citenamefont {Sipos}\ and\ \citenamefont
  {Goldenfeld}(2011)}]{Goldenfeld2011}%
  \BibitemOpen
  \bibfield  {author} {\bibinfo {author} {\bibfnamefont {M.}~\bibnamefont
  {Sipos}}\ and\ \bibinfo {author} {\bibfnamefont {N.}~\bibnamefont
  {Goldenfeld}},\ }\href@noop {} {\bibfield  {journal} {\bibinfo  {journal}
  {Phys. Rev. E}\ }\textbf {\bibinfo {volume} {84}},\ \bibinfo {pages} {035304}
  (\bibinfo {year} {2011})}\BibitemShut {NoStop}%
\bibitem [{\citenamefont {Shi}\ \emph {et~al.}(2015)\citenamefont {Shi},
  \citenamefont {Avila},\ and\ \citenamefont {Hof}}]{HofPreprint2015}%
  \BibitemOpen
  \bibfield  {author} {\bibinfo {author} {\bibfnamefont {L.}~\bibnamefont
  {Shi}}, \bibinfo {author} {\bibfnamefont {M.}~\bibnamefont {Avila}}, \ and\
  \bibinfo {author} {\bibfnamefont {B.}~\bibnamefont {Hof}},\ }\href@noop {}
  {\bibfield  {journal} {\bibinfo  {journal} {arXiv:1504.03304}\ } (\bibinfo
  {year} {2015})}\BibitemShut {NoStop}%
\bibitem [{\citenamefont {Shih}\ \emph {et~al.}(2015)\citenamefont {Shih},
  \citenamefont {Hsieh},\ and\ \citenamefont {Goldenfeld}}]{ShihPreprint15}%
  \BibitemOpen
  \bibfield  {author} {\bibinfo {author} {\bibfnamefont {H.-Y.}\ \bibnamefont
  {Shih}}, \bibinfo {author} {\bibfnamefont {T.-L.}\ \bibnamefont {Hsieh}}, \
  and\ \bibinfo {author} {\bibfnamefont {N.}~\bibnamefont {Goldenfeld}},\
  }\href@noop {} {\bibfield  {journal} {\bibinfo  {journal} {arXiv:1505.02807}\
  } (\bibinfo {year} {2015})}\BibitemShut {NoStop}%
\bibitem [{\citenamefont {Faisst}\ and\ \citenamefont
  {Eckhardt}(2003)}]{Faisst03}%
  \BibitemOpen
  \bibfield  {author} {\bibinfo {author} {\bibfnamefont {H.}~\bibnamefont
  {Faisst}}\ and\ \bibinfo {author} {\bibfnamefont {B.}~\bibnamefont
  {Eckhardt}},\ }\href@noop {} {\bibfield  {journal} {\bibinfo  {journal}
  {Phys. Rev. Lett.}\ }\textbf {\bibinfo {volume} {91}},\ \bibinfo {pages}
  {224502} (\bibinfo {year} {2003})}\BibitemShut {NoStop}%
\bibitem [{\citenamefont {Cvitanovi\'{c}}\ and\ \citenamefont
  {Gibbon}(2010)}]{Cvitanovic10}%
  \BibitemOpen
  \bibfield  {author} {\bibinfo {author} {\bibfnamefont {P.}~\bibnamefont
  {Cvitanovi\'{c}}}\ and\ \bibinfo {author} {\bibfnamefont {J.~F.}\
  \bibnamefont {Gibbon}},\ }\href@noop {} {\bibfield  {journal} {\bibinfo
  {journal} {Physica Scripta}\ }\textbf {\bibinfo {volume} {142}},\ \bibinfo
  {pages} {014007} (\bibinfo {year} {2010})}\BibitemShut {NoStop}%
\bibitem [{\citenamefont {{The data is publicly available}}()}]{data15b}%
  \BibitemOpen
  \bibfield  {author} {\bibinfo {author} {\bibnamefont {{The data is publicly
  available}}},\ }\href@noop {} {}\bibinfo {howpublished} {see
  \url{http://dx.doi.org/10.7488/ds/295}}\BibitemShut {NoStop}%
\end{thebibliography}%

\newpage
\newpage

\begin{center}
{\Large Supplemental Material}
\end{center}

\setcounter{footnote}{0}
\setcounter{equation}{0}
\setcounter{figure}{0}

\renewcommand{\thefigure}{S\arabic{figure}}
\renewcommand{\theequation}{S\arabic{equation}}

\subsection{Simulation parameters}

We perform direct numerical simulations of the incompressible Navier-Stokes equations, Eqs.(1) and (2) of the Main Text, with a forcing term using a fully de-aliased pseudospectral method \cite{Yoffe12} on a 3D periodic domain of length
$L_{box}=2\pi$. All simulations \cite{data15b} are performed with $32^3$ collocation points in a range of
kinematic viscosities $\nu$ changing from $0.1$ to $0.055$. This range corresponds to the
Taylor-Reynolds number $R_{\lambda}=2.61-4.72$ and the Reynolds number based on the box size
(as defined in the Main Text) $Re=53.8-97.82$; the product of the largest resolved wavenumber
$k_{max}$ and the Kolmogorov lengthscale $\eta$ is in the range of $k_{max}\eta=2.85-1.82$.
Simulations are evolved for $1271t_0$ time units, where the initial large-eddy turnover time
$t_0=L/U$, with $U$ and $L$ being the initial rms velocity and integral lengthscale, correspondingly.
For each Reynolds number we performed $100$ runs starting from random initial conditions, as discussed in the Main Text.

\subsection{Stability of the self-organized state}

\begin{figure}[!h]
 \begin{center}
   \includegraphics[width=\columnwidth]{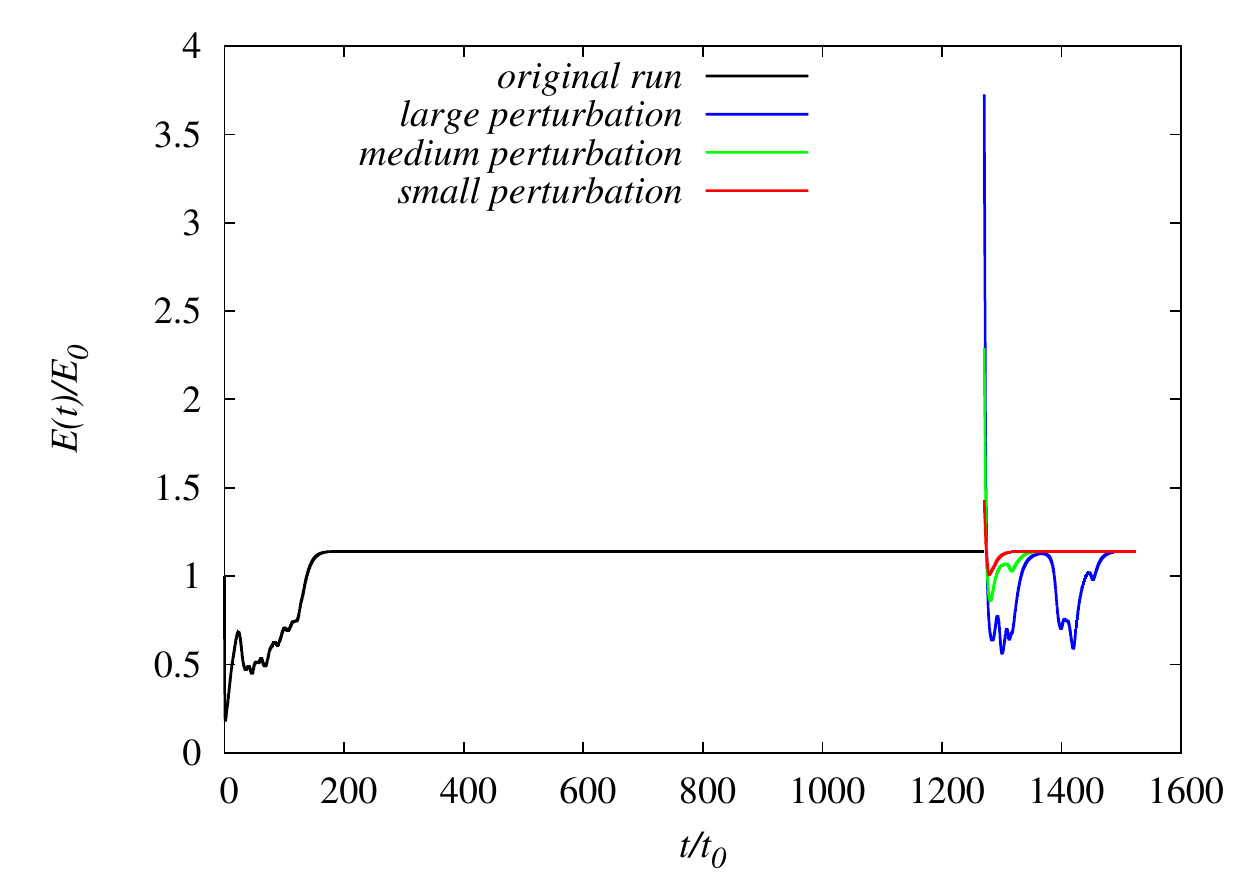}
 \end{center}
 \caption{Stability of the self-organized state for $Re=75.78$.}
 \label{fig:perturbations}
\end{figure}

Here we test the stability properties of the self-organized state. Below, we consider an example of $Re=75.78$. Simulations are started
from the self-organized state obtained at the end of a long run after a relaminarization event at the same Reynolds number. This state
is perturbed by a random initial perturbation of various amplitudes. In Fig.\ref{fig:perturbations} we plot the time evolution of
three runs, with relatively small, medium and large amplitudes. We observe that the small and medium amplitude
runs return to the self-organized state after some transient dynamics, while the large amplitude run becomes turbulent until its
dynamics again collapses to the self-organized state. The lifetime of turbulence in the large-amplitude run is similar but not equal to the
lifetime of the original run (shown in Fig.\ref{fig:perturbations}  for comparison), as can be expected from a process with an exponential survival probability (see Main Text).

\subsection{Survival Probabilities}

\begin{figure}[!h]
 \begin{center}
   \includegraphics[width=\columnwidth]{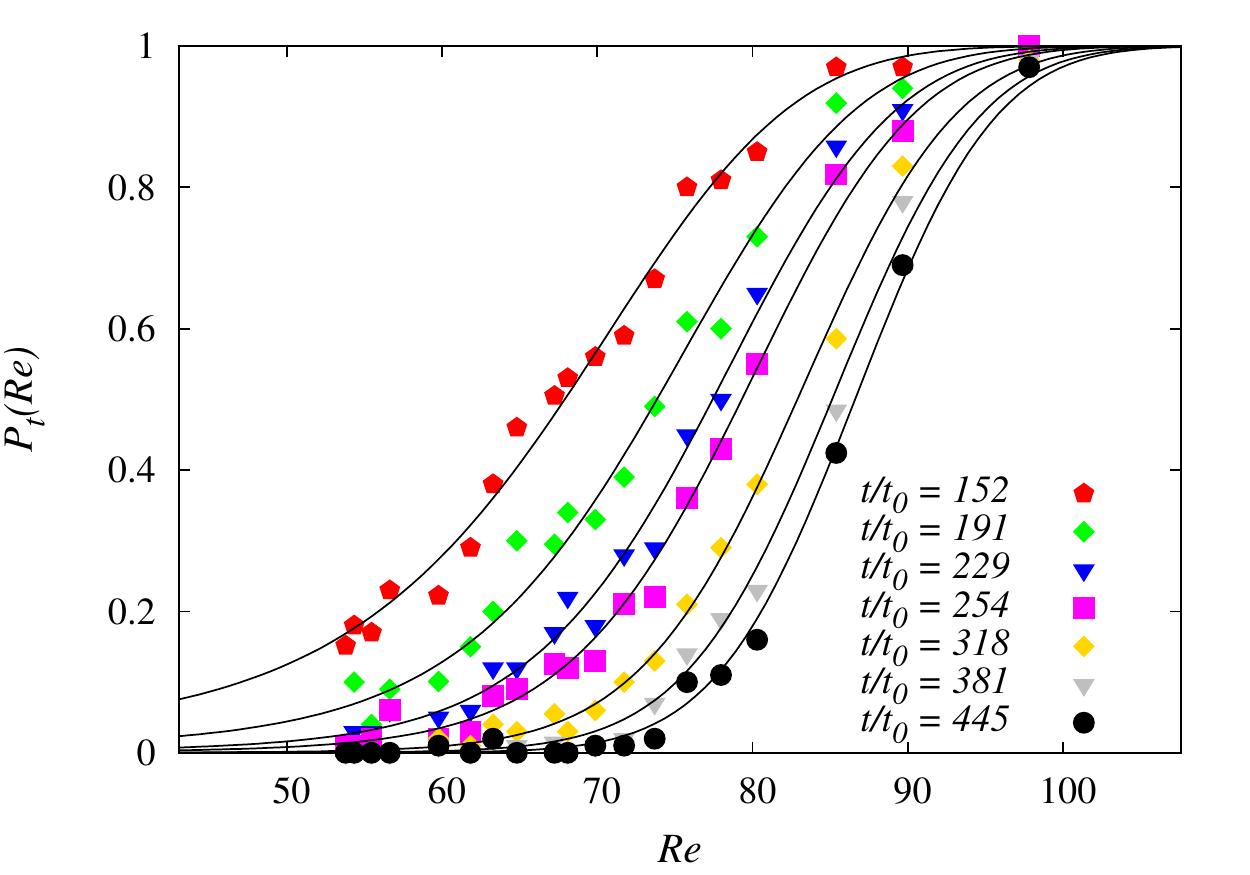}
 \end{center}
 \caption{Reynolds number dependence of the survival probabilities at different dimensionless observation times $t/t_0$.}
 \label{fig:survival_probSM}
\end{figure}

In order to further check that the statistics of relaminarization events follow a simple exponential law with a super-exponential lifetime, we combine the results obtained in the Main Text for the survival probability yielding
\begin{equation}
P(t)=\exp\left(-\frac{t-35}{20\exp\left[\exp(-3.48+\frac{Re}{19.23})\right]}\right).
\label{eq:probSM}
\end{equation}
We now use the collection of the relaminarization times for various
values of $Re$ from our simulations to calculate the survival probability
of the system being still turbulent after a fixed time $t$ as a
function of the Reynolds number. In Fig.~\ref{fig:survival_probSM} we
compare these results with the prediction of Eq.\eqref{eq:probSM} for
various values of the observation times $t$, where we held the
constant dividing $Re$ fixed while letting the additive constant
$a=3.48 \pm 0.51$ vary within its error bounds calculated from the
fitting procedure specified in the Main Text. We observe a good agreement
between the two data sets that provides further support to the exponential
form of the survival probability with a superexponential lifetime as
proposed in the Main Text.  The characteristic S-shape of the curves is
very similar to the results in wall-bounded flows \cite{Hof06,Hof08,Eckhardt08}.

\subsection{Simulations using a larger box size}
\begin{figure}[!h]
 \begin{center}
   \includegraphics[width=\columnwidth]{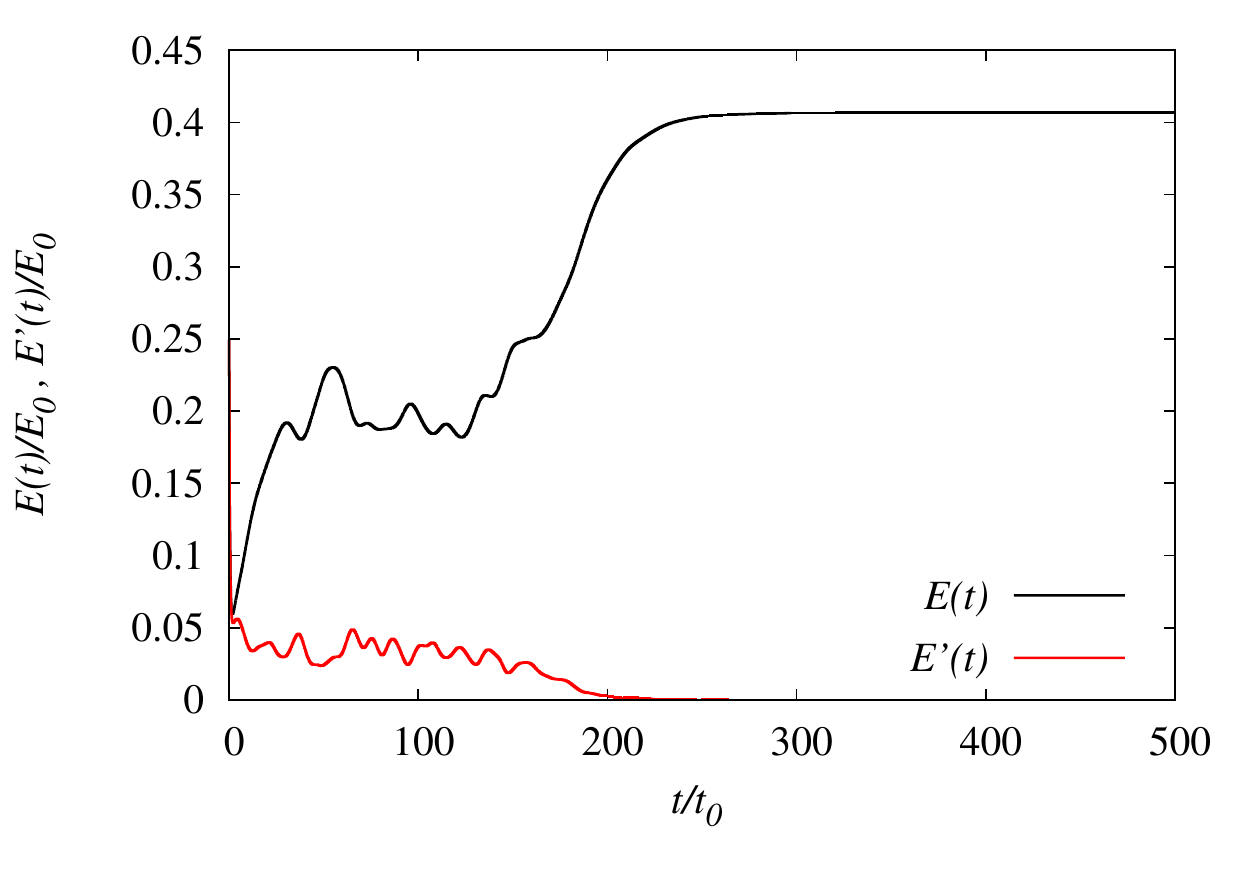}
 \end{center}
 \caption{Time evolution of the total energy $E(t)$ and the energy content of the small scales $E'(t)$
for $Re=68.10$ normalized by the initial energy $E_0$ for $L_{box} = 4 \pi$. Time is given in units of initial
large eddy turnover time $t_0=L/U$, where $U$ is the initial rms velocity and $L$ the initial integral scale.}
 \label{fig:largebox_run}
\end{figure}

\begin{figure}[!h]
 \begin{center}
   \includegraphics[width=\columnwidth]{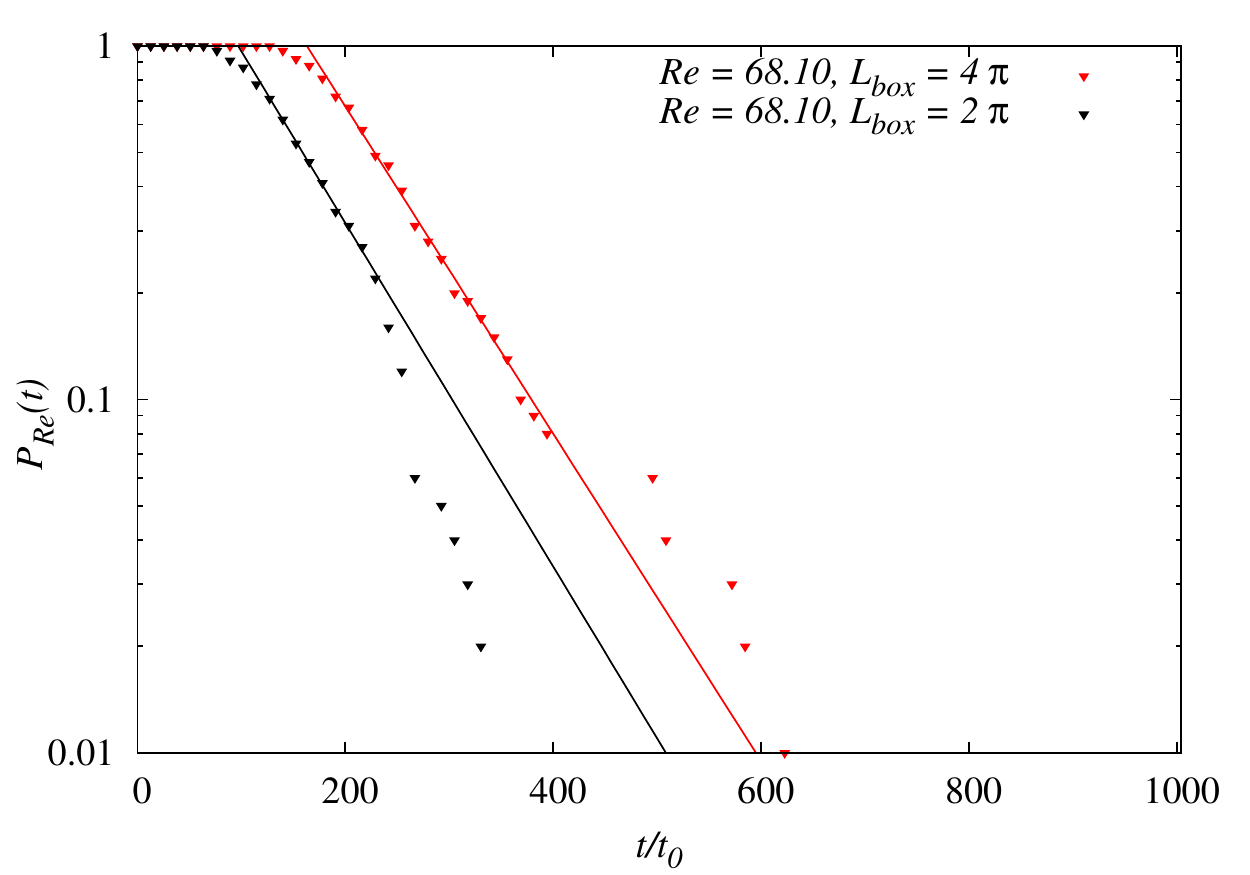}
 \end{center}
 \caption{
Survival probabilities as a function of the dimensionless time $t/t_0$
from the beginning of a simulation using
$L_{box} = 2 \pi$ (black) and $L_{box} = 4 \pi$ (red) for $Re=68.10$.
The solid lines represent fits of $P(t)=\exp(-t/\tau)$ to the respective
datasets. }
 \label{fig:largebox_prob}
\end{figure}

In order to demonstrate that the observed collapse of (transient) isotropic turbulence
is not an artefact of a too small simulation box ($L_{box}=2\pi$ in the runs presented in the Main Text),
an ensemble of 100 runs was created
using a box of size $L_{box}=4\pi$. Figure \ref{fig:largebox_run} shows the time evolution of the
total energy $E(t)$ and the energy content of the small scales $E'(t)$ for one simulation
belonging to this ensemble. The collapse of the small-scale
fluctuations is clearly visible, and the figure looks qualitatively very similar to
Fig.~1 of the Letter. The survival probabilities for this ensemble and an ensemble using $L_{box} = 2\pi$
at the same Reynolds number are shown in Fig.~\ref{fig:largebox_prob}, where we note that
the simulations using $L_{box} = 4\pi$ take longer to reach a (transient) turbulent
stationary state compared to the simulations using $L_{box} = 2\pi$. This is the reason
for the shift between the survival probabilities visible in the figure. However, the slopes of
the two exponentials, and, hence, the characteristic lifetimes obtained from the two datasets are almost indistinguishable, and certainly well within the error bars obtained from Eq.~\eqref{eq:probSM}.
 Since the simulations carried out on the larger box
size show very similar features to the data obtained from simulation using the
conventional box size $L_{box}=2\pi$, we conclude that the main results reported in the Letter are
not induced by the size of the simulation box.

\pagebreak 

\end{document}